\documentstyle[epsfig,12pt]{article}
\def\fun#1#2{\lower3.6pt\vbox{\baselineskip0pt\lineskip.9pt
\ialign{$\mathsurround=0pt#1\hfil##\hfil$\crcr#2\crcr\sim\crcr}}}

\newcommand{\be}{\begin{eqnarray}}
\newcommand{\ee}{\end{eqnarray}}

\begin{document}

\begin{figure}[htb]

\epsfxsize=6cm \epsfig{file=logo_INFN.epsf}

\end{figure}

\vspace{-4.75cm}

\Large{\rightline{Sezione ROMA III}}
\large{
\rightline{Via della Vasca Navale 84}
\rightline{I-00146 Roma, Italy}
}

\vspace{0.6cm}

\rightline{INFN-RM3 98/10}
\rightline{December 1998}

\normalsize{}

\vspace{0.5cm}

\begin{center}

{\Large \bf{A CONNECTION BETWEEN INCLUSIVE\\[0.1cm] SEMILEPTONIC
DECAYS OF\\[0.4cm] BOUND AND FREE HEAVY QUARKS}\footnote{\bf To
appear in Physical Review D.}}

\vspace{0.5cm}

\large{S.Ya. Kotkovsky$^{(a)}$, I.M. Narodetskii$^{(a)}$, S.
Simula$^{(b)}$ and K.A. Ter-Martirosyan$^{(a)}$}

\vspace{0.5cm}

\normalsize

$^{(a)}$Institute of Theoretical and Experimental Physics, Moscow,
117259 Russia\\[0.25cm]$^{(b)}$Istituto Nazionale di Fisica Nucleare,
Sezione Roma III\\ Via della Vasca Navale 84, I-00146 Roma (Italy)

\end{center}

\vspace{0.5cm}

\begin{abstract}

\noindent A relativistic constituent quark model, formulated on the
light-front, is used to derive a new parton approximation for the
inclusive semileptonic decay width of the $B$-meson. A simple
connection between the decay rate of a free heavy-quark and the one
of a heavy-quark bound in a meson or in a baryon is established. The
main features of the new approach are the treatment of the $b$-quark
as an on-mass-shell particle and the inclusion of the effects
arising from the $b$-quark transverse motion in the $B$-meson. In a
way conceptually similar to the deep-inelastic scattering case, the
$B$-meson inclusive width is expressed as the integral of the free
$b$-quark partial width multiplied by a bound-state factor related
to the $b$-quark distribution function in the $B$-meson. The
non-perturbative meson structure is described through various
quark-model wave functions, constructed via the Hamiltonian
light-front formalism using as input both relativized and
non-relativistic potential models. A link between spectroscopic
quark models and the $B$-meson decay physics is obtained in this
way. Our predictions for the $B \to X_c \ell \nu_{\ell}$ and $B \to
X_u \ell \nu_{\ell}$ decays are used to extract the $CKM$ parameters
$|V_{cb}|$ and $|V_{ub}|$ from available inclusive data. After
averaging over the various quark models adopted and including 
leading-order perturbative $QCD$ corrections, we obtain $|V_{cb}| =
(43.0 \pm 0.7_{exp} \pm 1.8_{th}) \cdot 10^{-3}$ and $|V_{ub}| =
(3.83 \pm 0.48_{exp} \pm 0.14_{th}) \cdot 10^{-3}$, implying
$|V_{ub} / V_{cb}| = 0.089 \pm 0.011_{exp} \pm 0.005_{th}$, in nice
agreement with existing predictions.

\end{abstract}

\newpage

\pagestyle{plain}

\section{Introduction}

\indent The investigation of inclusive semileptonic $B$-meson decays
can provide relevant information on the Cabibbo-Kobayashi-Maskawa
($CKM$) parameters $|V_{cb}|$ and $|V_{ub}|$ as well as on the
internal non-perturbative structure of the $B$-meson. In particular,
a precise knowledge of $V_{ub}$ is essential for the description of
the $CP$ violation within the Standard Model and indeed its
determination is one of the main goals of the beauty phenomenology.

\indent As far as the theoretical point of view is concerned, the
$QCD$-based operator product expansion ($OPE$) is widely recognized
as a consistent dynamical approach for investigating inclusive
heavy-flavour decays \cite{OPE}. It is also well known
\cite{DISTR,DSU96} that an adequate description of the end-point
region of the lepton spectrum requires a partial resummation of the
most singular terms of the $OPE$. In this way, the phenomenon of the
{\em Fermi motion} of the heavy quark inside the hadron, already
introduced into phenomenological models in a {\em ad hoc} way long
time ago \cite{ACM82}, emerges naturally in the $OPE$ approach. The
final result is conceptually similar to the leading-twist term of
deep-inelastic lepton-nucleon scattering case. In particular, the
effects of the Fermi motion are encoded in a heavy-quark
distribution function, whose first two moments can be expressed in
terms of the matrix elements of the heavy-quark  kinetic operator.
However, the first few moments do not exhaust the information hidden
in the heavy-quark distribution function, which cannot be calculated
yet from first principles; therefore, a particular functional form
is chosen in practice (cf. \cite {DISTR,DSU96}), introducing
consequently a model dependence. In this respect, the use of
phenomenological models, like the constituent quark model, could be
of great interest as a complementary approach to the $OPE$
resummation method.

\indent Till now, two main phenomenological approaches have been
applied to the description of the non-perturbative strong
interaction effects in inclusive heavy-flavour decays: the $ACCMM$
model \cite{ACM82} and the {\em exclusive} variant based on the
one-by-one summation of various final resonant channels
\cite{ISGW89}-\cite{LLSW97}. The impact of the Fermi motion in the
parton model has also been addressed in Refs.
\cite{JPP94}-\cite{GNST96}, where the effects due to the internal
motion of the $b$-quark inside the $B$-meson have been encoded in
a model-dependent quark distribution function. In Ref. \cite{JPP94}
the latter has been related to the fragmentation function of
heavy-quarks into heavy mesons, while in Refs. \cite{MTM96,GNST96}
the bound-state effects have been incorporated via the following
distribution function
 \be
    \label{1.1}
    F(x) \equiv \int d\vec{p}_{\perp} ~ |\psi(x, p_{\perp}^2)|^2,
 \ee
where $|\psi(x, p_{\perp}^2)|^2$ is the probability to find the
$b$-quark carrying a light-front ($LF$) fraction $x = p^+_b /
P^+_B$ of the $B$-meson momentum and a transverse relative
momentum squared $p_{\perp}^2$ ($\equiv |\vec{p}_{\perp}|^2$). In
Ref. \cite{GNST96} the model wave function $\psi(x, p_{\perp}^2)$
has been constructed via the Hamiltonian $LF$ formalism (cf. Refs.
\cite{LIGHT-FRONT,SIM96}) using as input the canonical wave
functions corresponding to various quark potential models.

\indent The important feature of the approach of Refs.
\cite{MTM96,GNST96} is the treatment of the $b$-quark as a virtual
particle with mass $m_b^2 = x^2 M_B^2$, where $M_B$ is the
$B$-meson mass, while the effects due to the transverse motion of
the $b$-quark are neglected everywhere except in the calculation of
the distribution function $F(x)$ in Eq. (\ref{1.1}). The final
expressions obtained for the semileptonic branching ratios and the
lepton energy distributions clearly exhibit a close analogy with
the deep-inelastic lepton-nucleon scattering case.

\indent The aim of this paper is to generalize the work of
Refs. \cite{JPP94}-\cite{GNST96} by developing a more refined
expression for the inclusive semileptonic decay width\footnote{A
preview of our approach can be found in \cite{K97}.}. The new
features are the treatment of the $b$-quark as an on-mass-shell
particle with mass $m_b$ (as it is required in the Hamiltonian
$LF$ formalism) and the full inclusion of the effects due to the
$b$-quark transverse momenta. Our main result is the derivation of
a new parton formula for the differential inclusive width, which
is similar to the one derived by Bjorken {\it et al.} \cite{BDT92}
in case of infinitely heavy $b$- and $c$-quarks, viz.
 \be
    \label{1.2}
    {d\Gamma_{SL} \over dq^2d q_0} = {d\Gamma^{(free)}_{SL} \over
    dq^2} ~ \omega(q^2,q_0),
 \ee
where $d\Gamma^{(free)}_{SL} / dq^2$ is the free-quark
differential decay rate and the function $\omega(q^2, q_0)$
incorporates the effects of the Fermi motion of the heavy quark
expressed in terms of the $b$-quark distribution function
$|\psi(x, p_{\perp}^2)|^2$. In Eq. (\ref{1.2}) $q = p_{\ell} +
p_{\nu_{\ell}}$ is the four-momentum of the lepton pair, and $q_0$
is the dilepton energy in the $B$-meson rest frame. The structure
of Eq. (\ref{1.2}) suggests that in the limit of heavy quarks with
infinite mass (i.e., $m_b \to \infty$ and $m_c \to \infty$) one has
 \be
    \label{1.3}
     \int dq_0 ~ \omega(q^2, q_0) = 1.
 \ee
which means that the total inclusive width of the hadron is the
same as the total inclusive width at the free quark level. The
corrections to the free-quark decay picture are mainly due to the
difference between the quark mass $m_b$ and the meson mass $M_B$
as well as to the {\em primordial} motion of the $b$-quark inside
the $B$-meson. These non-perturbative corrections vanish in the
heavy-quark limit $m_b \to \infty$, but at finite values of the
$b$-quark mass a new parton description of inclusive semileptonic
decays, based on the constituent quark model, is derived.

\indent The plan of the paper is as follows. Section 2 contains a
brief discussion of the kinematics relevant to inclusive
semileptonic $B$-meson decays. In Section 3 we derive our main
result, Eq. (\ref{1.2}), for the differential inclusive
semileptonic decay rate. In Section 4 we compute the semileptonic
branching ratio for the processes $B \to X_c \ell \nu_{\ell}$ and
$B \to X_u \ell \nu_{\ell}$, paying particular attention to the
extraction of $|V_{cb}|$ and $|V_{ub}|$ from available
inclusive data. The dependence upon the quark model parameters will
be estimated through the use of different meson wave functions,
either obtained in a phenomenological way (as in Ref. \cite{MTM96})
or constructed via the Hamiltonian $LF$ formalism from quark
potential models (as in Ref. \cite{GNST96}). Finally,
our conclusions are summarised in Section 5.

\section{Kinematics}

\indent The inclusive semileptonic width $\Gamma_{SL}$ for the
decay process $B \to X_{q'} \ell \bar{\nu}_{\ell}$, where $\ell =
e, \mu$ or $\tau$ and $X_{q'}$ is any possible hadronic state
containing a charm quark $(q' = c)$ or a light quark $(q' = u)$,
can be written in terms of the contraction among the leptonic
tensor $L^{\alpha \beta}$ and the hadronic one $W_{\alpha \beta}$
\cite{OPE}:
 \be
    \label{2.1}
    \Gamma_{SL} = {1 \over (2\pi)^3} {G_F^2|V_{bq'}|^2 \over
    M_B} \int d^4q \int d\tau_{\ell} ~ L^{\alpha \beta} W_{\alpha
    \beta},
 \ee
where $d^4q = 2\pi |\vec{q}| dq^2 dq_0$, $d\tau_{\ell} =
|\vec{p}_{\ell}| ~ d\Omega_{\ell} / (16 \pi^2 ~ \sqrt{q^2})$ is the
leptonic phase space, $d\Omega_{\ell}$ is the solid angle of the
charged lepton $\ell$, $|\vec{p}_{\ell}| = \sqrt{q^2} ~ \Phi_{\ell}
/ 2$ is its momentum in the dilepton center-of-mass frame and
$\Phi_{\ell} \equiv \sqrt{1 - 2\lambda_+ + \lambda^2_-}$, with
$\lambda_{\pm} \equiv (m^2_{\ell} \pm m^2_{\nu_{\ell}}) / q^2$.
The tensors $L^{\alpha \beta}$ and $W_{\alpha \beta}$ in Eq.
(\ref{2.1}) are explicitly given by
 \be
    \label{2.4}
    L^{\alpha \beta} =  2 [p_{\ell}^{\alpha} p_{\nu_{\ell}}^{\beta}
    + p_{\ell}^{\beta} p_{\nu_{\ell}}^{\alpha} - g^{\alpha \beta}
    (p_{\ell} \cdot p_{\nu_{\ell}}) + i\epsilon^{\alpha \beta
    \gamma \delta} p_{\ell \gamma} p_{\nu_{\ell} \delta}],
 \ee
 \be
    \label{2.5}
    W_{\alpha \beta} = (2\pi)^3 \sum\limits_n \int
    \prod\limits_{i=1}^n {d\vec{p}_i \over (2 \pi)^3 2E_i} ~
    \delta^4(P_B - q - \sum\limits_{i=1}^n p_i) ~
    <B| j^+_{\alpha}(0) |n> ~ <n| j_{\beta}(0) |B>  ,
 \ee
respectively. The summation in Eq. (\ref{2.5}) includes all possible
final hadronic states, $j_{\alpha}(0)$ is the weak current mediating
the decay $b \to q'$ and $P_B$ is the $B$-meson four-momentum. The
hadronic tensor (\ref{2.5}) is function of the two invariants $q^2 =
q \cdot q$ and $q_0 \equiv (q \cdot P_B) / M_B$, where the latter is
related to the invariant mass $M_X$ of the final hadronic system by:
$q_0 = (M_B^2 + q^2 - M_X^2) / 2M_B$. In what follows we will
consider the $B$-meson rest frame.

\indent The integral over the leptonic phase space in Eq.
(\ref{2.1}) is given by
  \be
     \label{2.6}
     \int d\tau_{\ell} ~  L^{\alpha \beta} = {1 \over 4\pi}
     {|\vec{p}_{\ell}| \over \sqrt{q^2}} <L^{\alpha \beta}>, 
 \ee
with
 \be
    \label{2.7}
    <L^{\alpha \beta}> = {1 \over 4\pi} \int d\Omega_{\ell} ~
    L^{\alpha \beta} = {2 \over 3} \left \{ (1 + \lambda_1)
    (q^{\alpha} q^{\beta}- g^{\alpha\beta} q^2) + {3 \over 2}
    \lambda_2 g^{\alpha \beta} q^2 \right \}, 
 \ee
where $\lambda_1 \equiv \lambda_+ - 2\lambda^2_-$ and $\lambda_2
\equiv \lambda_+ - \lambda^2_-$.

\indent Introducing the dimensionless kinematical variables $t
\equiv q^2 / m_b^2$ and $s \equiv M_X^2 / m_b^2$, the semileptonic
width (\ref{2.1}) can be cast into the form
 \be
    \label{2.8}
    \Gamma_{SL} = {G_F^2 m_b^5 \over (4\pi)^3} ~ |V_{bq'}|^2 ~ 
    \int\limits_{t_{min}}^{t_{max}} dt ~ \Phi_{\ell}(t)
    \int\limits_{s_{min}}^{s_{max}} ds ~ {|\vec{q}| \over m_b}
    {<L^{\alpha \beta}> W_{\alpha \beta} \over M_B^2},
 \ee
where
 \be
    \label{2.9}
    {2 |\vec{q}| \over m_b} \equiv \alpha(t, s) = {1 \over x_0}
    \sqrt{(1 + x_0^2 t - x_0^2 s)^2- 4 x_0^2 t}.
 \ee
and $x_0 \equiv m_b / M_B$. In Eq. (\ref{2.8}) the limits of
integrations in the $t$-$s$ plane are given by: $s_{min} = \zeta^2
/ x_0^2$, $s_{max} = (1 - x_0 \sqrt{t})^2 / x_0^2$, $t_{min} =
m_{\ell}^2 / m_b^2$ and $t_{max} = (1 - \zeta)^2 / x_0^2$, where
$\zeta \equiv M_{thr} / M_B$,  with $M_{thr}$ being the lowest mass
of the final hadronic state ($M_{thr} = M_D$ in case of the $b \to
c$ transition and $M_{thr} = M_{\pi}$ for the $b \to u$ transition).

Before closing this section we note that the right-hand side of Eq.
(\ref{2.8}) is expressed in terms of an integral over physical
spectral densities and phase space, both depending on meson masses.
As it will be shown in the next Section, in our parton picture the
heavy-quark mass $m_b$ emerges as the relevant parameter.

\section{Light front constituent quark model approximation for
$W_{\mu \nu}$}

\indent In this section we apply the constituent quark model to the
treatment of semileptonic beauty decays, $B \to X_q \ell \nu_{\ell}$,
in close analogy with the parton approximation in deep-inelastic
lepton-nucleon scattering. Our approach is based on ~ i) the
hypothesis of quark-hadron duality, which assumes that, when a
sufficient number of exclusive hadronic decay modes is summed up,
the decay probability into hadrons equals the one of its partons,
and ~ ii) the dominance of the valence component in the $B$-meson
wave function. Following this assumption, the hadronic tensor
$W_{\alpha\beta}$ is given through the optical theorem by the
imaginary part of the quark box diagram  describing the forward
scattering amplitude \cite{JPP94}-\cite{GNST96}:
 \be
    \label{3.1}
     W_{\alpha\beta} = \int\limits_0^1 {dx \over {x}} \int
     d\vec{p}_{\perp} ~ w_{\alpha \beta}^{(bq')}(p_b, p_{q'}) ~
     \delta[(p_b - q)^2 - m_{q'}^2] ~ \theta(\varepsilon_{q'}) ~
     |\psi(x, p_{\perp}^2)|^2,
 \ee
where $p_b \equiv (p_b^+, p_b^-, \vec{p}_{\perp}) = (x M_B,
{m_b^2 + p_{\perp}^2 \over x M_B}, \vec{p}_{\perp})$ with
$p_b^2 = p_b^- p_b^+ - p_{\perp}^2 = m_b^2$ and the quark tensor
$w_{\alpha \beta}^{(bq')}(p_b, p_{q'})$ is defined analogously to
the lepton tensor in Eq. (\ref{2.4}):
 \be
    \label{3.2}
    w_{\alpha \beta}^{(bq')}(p_b, p_{q'}) = 4 [p_{q' \alpha} p_{b
    \beta} + p_{q' \beta} p_ {b \alpha} - g_{\alpha \beta}(p_{q'}
    \cdot p_b) + i\epsilon_{\alpha \beta \gamma \delta}
    p_{q'}^{\gamma} p_b^{\delta}]
 \ee
Equation (\ref{3.1}) corresponds to the average of the free-quark
decay distribution over the motion of the heavy quark, described
by the distribution function $|\psi(x, p_{\perp}^2)|^2$, whose
normalization is given by $\int\limits_0^1 dx \int
d\vec{p}_{\perp} ~ |\psi(x, p_{\perp}^2)|^2 = 1$. In Eq.
(\ref{3.1}) the function $\theta(\varepsilon_{q'})$, where
$\varepsilon_{q'}$ is the $q'$-quark energy, is inserted for
consistency with the use of the valence $LF$ wave function
$\psi(x, p_{\perp}^2)$, while the $\delta$-function, expressing the
decay of the $b$-quark to a $q'$-quark, can be rewritten as
 \be
    \label{3.3}
    \delta[(p_b - q)^2 - m_{q'}^2] = {x \over x_0} {m_b \over q^+}
    ~ \delta(p_{\perp}^2 - p_{\perp}^{*2})
 \ee
where $q^+$ is the $LF$ {\em plus} component of the dilepton
momentum, $q^+ = q_0 + |\vec{q}|$, and
 \be
    \label{3.4}
    p_{\perp}^{*2} = m^2_b \left[{x \over x_0} {m_b \over q^+} (1 +
    t - \rho) - ({x \over x_0} {m_b \over q^+})^2 t - 1 \right],
 \ee
with $\rho \equiv m_{q'}^2 / m_b^2$ being the quark mass ratio
squared.

\indent We now substitute Eq. (\ref{3.1}) into (\ref{2.8}) and use
the fact that the contraction of the {\it averaged} lepton tensor
$<L^{\alpha \beta}>$ and the quark tensor $w^{bq'}_{\alpha \beta}$
does not depend on $x$ and $s$ and, therefore, can be taken out of
the integral over $x$ in Eq. (\ref{3.1}); the explicit expression
for the tensor contraction is given in the Appendix (see Eqs.
(\ref{A.6}-\ref{A.9})). The differential decay rate becomes
 \be
    \label{3.6}
    {d^2\Gamma_{SL} \over dt ds} = {G_F^2 m_b^5 \over (4\pi)^3}
    ~ |V_{bq'}|^2 ~ {|\vec{q}|\over m_b} ~ \Phi_{\ell}(t) ~
    {<L^{\alpha \beta}> w^{(bq')}_{\alpha \beta} \over M_B^2} ~
    {\pi \over x_0} {m_b \over q^+} ~ \int_{x_1}^{min[1,x_2]} dx ~
    |\psi(x, p_{\perp}^{*2})|^2, 
 \ee
where the integration limits follow from the condition
$p_{\perp}^{*2} \ge 0$, viz.
 \be
    \label{3.7}
    x_{1, 2} = x_0 {q^+ \over \tilde{q}^{\pm}} = x_0 {q_0 +
    |\vec{q}| \over \tilde{q}_0 \pm  |\vec{\tilde{q}}|}
 \ee
with $\tilde{q}_0$ ($\vec{\tilde{q}}$) being the energy
(three-momentum) of the lepton pair in the $b$-quark rest frame,
viz.
 \be
    \label{3.8}
    {2 \tilde{q}_0 \over m_b} = 1 + t - \rho, ~~~~~~~~	
    {2 |\vec{\tilde{q}}| \over m_b} \equiv \tilde{\alpha}(t, \rho)
    = \sqrt{(1 + t - \rho)^2 - 4t}. 
 \ee
To proceed further we do not need the explicit expression for
the contraction $<L^{\alpha \beta}> w^{(bq')}_{\alpha \beta}$,
which any way can be easily calculated using Eqs. (\ref{2.7}) and
(\ref{3.2}) (see Eqs. (\ref{A.6}) and (\ref{A.9}) in the Appendix).
Instead we note that the same contraction appears in the
differential semileptonic decay width of a free $b$-quark into a
$q'$-quark
 \be
    \label{3.9}
    {d\Gamma^{(free)}_{SL}(t) \over dt} = {G_F^2 m_b^5 \over
    (4\pi)^3} ~ |V_{bq'}|^2 ~ {|\vec{\tilde{q}}|  \over m_b} ~ 
    \Phi_{\ell}(t) ~ {<L^{\alpha \beta}> w^{(bq')}_{\alpha \beta}
    \over m_b^4},
 \ee

\indent Using Eq. (\ref{3.9}) we can express the contraction
$<L^{\alpha \beta}> w^{bq'}_{\alpha \beta}$ through
$d\Gamma^{(free)}_{SL} / dt$. Then, from Eq. (\ref{3.6}) we obtain 
our main result for the semileptonic decay width of the $B$-meson
 \be
    \label{3.11}
    {d\Gamma_{SL} \over dt} = {d\Gamma^{(free)}_{SL}(t) \over dt}
    \int_{s_{min}}^{s_{max}} ds ~ \omega(t, s),
 \ee
where the bound-state function $\omega(t, s)$ is defined as
 \be
    \label{3.12}
    \omega(t, s) = m_b^2 x_0 ~ {\pi m_b \over q^+} ~ {|\vec{q}|
    \over |\vec{\tilde{q}}|} ~ \int\limits_{x_1}^{min[1, x_2]} dx ~
    |\psi(x, p_{\perp}^{*2})|^2 
\ee
In Eq. (\ref{3.11}) the region of integration over the final
invariant hadron mass is characterized by a {\em quark} threshold
$M_{thr}^{(0)} = m_c$, defined through the condition $x_1 = min[1,
x_2]$ in Eq. (\ref{3.12}), which differs from the hadronic threshold
(i.e., $M_D$ for $q' = c$ or $M_{\pi}$ for $q' = u$)\footnote{Note
that the {\em quark} threshold $M_{thr}^{(0)} =  m_c$ differs also
from the parameter $M_{th}$, which was introduced  in Refs.
\cite{MTM96,GNST96} with the aim of separating the exclusive $D$ and
$D^*$ channels from the hadron continuum.}. Explicitly one  gets:
$s_{min} = \rho = (m_c / m_b)^2$ and $s_{max} = x_0^2 (1 - x_0 
\sqrt{t})^2$. Moreover, since the inequality $t \le (1 -
\sqrt{\rho})^2$ follows from Eq. (\ref{3.8}), the bound-state factor
(\ref{3.12}) as well as the structure functions (\ref{A.8}) are
identically vanishing for $q^2 \ge (m_b - m_{q'})^2$. 

\indent To sum up, the inclusive semileptonic decay width, Eq.
(\ref{3.11}), is expressed as an integral of two factors. The first
one is the {\em parton} differential decay rate (\ref{3.9}) of a
free heavy quark, which sets the overall scale for the decay rates
and does not depend on the spectator quark. The second factor (Eq.
(\ref{3.12})), which incorporates the non-perturbative  corrections
to the free-quark result (now depending on the spectator quark), is
given by an integral over the distribution function of the
$b$-quark. In the heavy-quark limit, $m_b \to \infty$, the $b$-quark
distribution in the $B$-meson becomes a delta function peaked at $x
= 1$, (more precisely $\delta(x - 1) \cdot \delta(\vec{p}_{\perp})$),
which implies that the function $\omega(t, s)$ goes to $\delta(s -
\rho)$ for $m_b \to \infty$, leading to the sum rule (\ref{1.3}). At
finite values of the  $b$-quark mass our result for $\Gamma_{SL}$
exhibits an  $m_b$-dependence of the following general form:
$\Gamma_{SL}  \propto m_b^5 [1 + c / m_b + O(1 / m_b^2)]$, where $c$
is a  non-vanishing coefficient depending on the particular quark 
model adopted. However, the mass $m_b$ is the constituent mass  of
the $b$-quark, which may differ from the pole quark mass $\mu_b$
commonly appearing in the $OPE$ of the $B$-meson decay rate (see
Ref. \cite{OPE} and also Ref. \cite{SIM98}). Assuming $m_b = \mu_b 
- c/5 + O(1 / \mu_b)$, the well-known result \cite{OPE} of the
absence  of the $1 / \mu_b$ corrections to the free-quark decay may
be  recovered. The above argument is completely analogous to that
used to  eliminate the $1 / \mu_b$ corrections from the total width
in the  $ACCMM$ model \cite{CBB}.    

\section{Numerical results}

\indent After having described the theoretical tools involved in
the calculation of the inclusive rate, we now focus on the practical
way for the extraction of the $CKM$ parameters $|V_{cb}|$ and
$|V_{ub}|$ from available inclusive data.

\indent The non-perturbative ingredient in Eq. (\ref{3.12}) is the
meson wave function $\psi(x, p_{\perp}^2)$. In what follows, we
will adopt both a phenomenological ans\"atz and the wave functions
corresponding to various quark potential models. As for the
phenomenological wave function, we use the exponential ans\"atz
introduced in Ref. \cite{MTM96}, which reads as
 \be
    \label{MTM}
     \psi(x, p_{\perp}^2) = {{\cal{N}} \over \sqrt{(1 - x)}} ~
     exp \left[-{\lambda_0 \over 2} \left( {1 - x \over \xi_0} +
     {\xi_0 \over 1 - x} ~ (1 + {p_{\perp}^2 \over m_{sp}^2})
     \right) \right],
 \ee
where $m_{sp}$ is the mass of the spectator quark in the $B$-meson,
$\xi_0 \equiv m_{sp} / M_B$, ${\cal{N}}$ is a normalization
constant and $\lambda_0 \simeq 1$ is an adjustable parameter
obtained in Ref. \cite{MTM96} from a fit of the experimental data
on the differential decay rate of the exclusive process $B \to D^*
\ell \nu_{\ell}$. In what follows we will refer to the
phenomenological ans\"atz (\ref{MTM}) as the quark model $A$. The
other models considered ($B$ to $E$) are based on $LF$-type meson
wave functions, which can be written in terms of the radial wave
function $w(p^2)$ corresponding to a particular quark potential
model as \cite{GNST96}
 \be
    \label{4.1}
    \psi^{LF}(x, p_{\perp}^2) = \sqrt{{M_0 \over 4x(1 - x)} \left[
    1 - \left( {m_b^2 - m_{sp}^2 \over M_0^2} \right)^2
    \right]} ~ {w(p^2) \over \sqrt{4\pi}},
 \ee
where $M_0 \equiv \sqrt{m_b^2 + p^2} + \sqrt{m_{sp}^2 + p^2}
= \sqrt{(p_{\perp}^2 + m_b^2) / x + (p_{\perp}^2 + m_{sp}^2) / (1 -
x)}$ is the free mass and $p^2 \equiv p_{\perp}^2 + p_z^2$, with
$p_z = (x - 1/2) M_0 + (m_{sp}^2-m_b^2) / 2M_0$. In particular,
models $B - E$ correspond to the radial wave functions $w(p^2)$
of the potential models of Refs. \cite{F97}, \cite{IS95},
\cite{NCS92} and \cite{GI85}, respectively. The main difference
among the various quark models lies in the behaviour of $w(p^2)$
at high values of the internal momentum $p$. Models $B$ and $C$
correspond to a soft Gaussian ans\"atz, which takes into account
mainly the effects of the confinement scale, whereas the functions
$w(p^2)$ corresponding to models $D$ and $E$ exhibit high momentum
components generated by the effective one-gluon exchange part of
the interquark potential. In case of models $A$, $D$ and $E$ the
distribution function $F(x)$ (Eq. (\ref{1.1})) has been already
calculated in Ref. \cite{GNST96}. It turns out that the models $A
- D$ yield quite similar results for $F(x)$, whereas model $E$
predicts a remarkably broader $x$-distribution. In terms of the
mean value $\langle x \rangle \equiv \int_0^1 dx ~ x ~ F(x)$ and
the variance $\sigma^2 \equiv \int_0^1 dx ~ (x - \langle x
\rangle)^2 ~ F(x)$, one gets $\langle x \rangle \simeq 0.90$ and
$\sqrt{\sigma^2} \simeq 0.06$ for models $A - D$, while in case of
model $E$ (the Godfrey-Isgur ($GI$) potential model) one has
$\langle x \rangle \simeq 0.87$ and $\sqrt{\sigma^2} \simeq
0.09$\footnote{Note that $\langle x \rangle$ does not generally
coincide with the location of the maximum of $F(x)$ as well as with
the value of $x_0$.}. Such a striking difference is directly related
to the larger mean value of the internal momentum characterizing
the $GI$ model with respect to the others cases $A - D$. The
values of the constituent quark masses as well as the mean value
$<p^2> \equiv \int_0^{\infty} dp ~ p^4 ~ |w(p^2)|^2$ for the
different quark models considered in this paper are collected in
Table 1.

\indent We have calculated Eqs. (\ref{3.11}-\ref{3.12}) in case of
the decay procesess $B \to X_c \ell \nu_{\ell}$ and $B \to X_u
\ell \nu_{\ell}$ adopting the five quark models $A$ to $E$ (in all
these models the physical mass of the $B$-meson, $M_B = 5.279 ~
GeV$, has been taken from the recent $PDG$ publication \cite{PDG96}).
We have also considered that perturbative $QCD$  corrections lead
approximately to an additional multiplicative factor $J^{pert}_{LO}$
in Eq. (\ref{3.11}), which typically reduces the semileptonic decay
width. At leading order the correction associated with the running
coupling constant $\alpha_s$ is well known \cite{pQCD} and for the
$b \to c$ ($b \to u$) transition we  will consider in what follows
the value $J^{pert}_{LO} =  0.90$ ($0.85$). Our results for the
branching ratio $\Gamma_{SL} / \Gamma_B^{(exp)}$, where
$\Gamma_B^{(exp)}$ is the experimental $B$-meson width
($\Gamma_B^{(exp)} = 1 / \tau_B^{(exp)}$ with $\tau_B^{(exp)} = 1.57
\pm 0.04 ~ ps$ \cite{BHP96}), are collected in Table 2 for the
process $B \to X_c \ell \nu_{\ell}$ with the branching ratio being
given in units of $10^{-2} \cdot (|V_{cb}| / 0.040)^2 \cdot
(\tau_B^{(exp)} / 1.57 ~ ps)$. It can be seen that the
non-perturbative effects, mocked up in the $b$-quark distribution
function $|\psi(x, p_{\perp}^2)|^2$, are mainly related  both to the
broadening of the $x$-distribution, generated by the high-momentum
components of the $B$-meson wave function (see model $E$), and to
the quark mass ratio $\sqrt{\rho} = m_{q'} / m_b$ (see Table 1). The
values predicted for the semileptonic branching ratio $\Gamma_{SL} /
\Gamma_B^{(exp)}$ exhibit a model dependence of about $\pm 10 \%$.
Finally, assuming the experimental world average value
$Br_{SL}^{(exp)}(B \to X_c e \nu_e) = (10.43 \pm 0.24) \%$
\cite{PDG96}, the $CKM$ matrix element $|V_{cb}|$ can easily be
obtained from the predicted values of the $SL$ branghing ratio and
the corresponding results are reported in Table 2. The average over
the various quark-model predictions yields
 \be
    \label{4.2}
    |V_{cb}| = (43.0 \pm 0.7_{exp} \pm 1.8_{th}) \cdot 10^{-3}
    \cdot \sqrt{{Br_{SL}^{(exp)} \over 10.43 ~ \%}} \cdot
    \sqrt{{1.57 ~ ps \over \tau_B^{(exp)}}},
 \ee
where the experimental errors of the branching ratio and the
$B$-meson lifetime have been taken in quadrature. Our $LF$
prediction (\ref{4.2}) is consistent with the updated
"experimental" determination of $|V_{cb}|$
\cite{SKWA96}\footnote{The value quoted in Ref. \cite{SKWA96}
corresponds to $Br_{SL}^{(exp)} = 10.77 \pm 0.43 \%$ and
$\tau_B^{(exp)} = 1.60 \pm 0.03 ~ ps$. After correcting for
the values adopted in this paper, one gets $|V_{cb}|_{incl} =
(39.5 \pm 0.9_{exp} \pm 4.0_{th}) \cdot 10^{-3}$.},
$|V_{cb}|_{incl} = (39.8 \pm 0.9_{exp} \pm 4.0_{th}) \cdot
10^{-3}$, as well as with the recent $OPE$ analysis of Ref.
\cite{BI96}, $|V_{cb}|_{incl} = (41.3 \pm 1.6_{exp} \pm 2.0_{th})
\cdot 10^{-3}$.

\indent The other inclusive process we want to consider is the
semileptonic decay $B \to X_u \ell \nu_{\ell}$. The existence of
the  $b \to u \ell \nu_{\ell}$ transition has been demonstrated
few years ago by the $CLEO$ \cite {CLEO90,CLEO93} and $ARGUS$
\cite{ARGUS90} collaborations through the observation of semileptonic
$B$-meson decays with leptons that are too energetic to originate
from the $b \to c \ell \nu_{\ell}$ transition. Very recently, the
$CLEO$ collaboration \cite{CLEO95} has reported the first signal for
exclusive semileptonic decays of the $B$-meson into charmless final
states. In Ref. \cite{ALEPH} the $ALEPH$ collaboration has announced
a model-independent measurement of the inclusive $b \to u \ell
\nu_{\ell}$ width, viz. $Br(b \to u \ell \nu_{\ell}) = (0.16 \pm
0.04) ~ \%$. The most important application of the analysis of the
inclusive decay $B \to X_u \ell \nu_{\ell}$ is the extraction of the
$CKM$ parameter $|V_{ub}|$. Our results obtained for the branching
ratio $\Gamma_{SL} / \Gamma_B^{(exp)}$ and for the $CKM$ parameter,
$|V_{ub}|$, extracted adopting the $ALEPH$ value for the
experimental semileptonic branching ratio (i.e, $Br_{SL}^{(exp)}(B
\to X_u e \nu_e) = (0.16 \pm 0.04) \%$), are reported in Table 3.
Using the values of $|V_{cb}|$ obtained in Table 2, also our
predictions for the ratio $|V_{ub} / V_{cb}|$ are reported in Table
3. It can be seen that the model dependence of the predicted
branching ratio is about $\pm 10 \%$ as in case of the decay process
$B \to X_c \ell \nu_{\ell}$ (see Table 2). Averaging our predictions
over the various quark models, one gets
 \be
    \label{4.3}
    |V_{ub}| & = & (3.83 \pm 0.48_{exp} \pm 0.14_{th}) \cdot
    10^{-3} \cdot \sqrt{{Br_{SL}^{(exp)} \over 0.16 ~ \%}} \cdot
    \sqrt{{1.57 ~ ps \over \tau_B^{(exp)}}} \\
    \label{4.4}
    |V_{ub} / V_{cb}| & = & 0.089 \pm 0.011_{exp} \pm 0.005_{th}.
 \ee
Our result for $|V_{ub} / V_{cb}|$ (Eq. (\ref{4.4})) is consistent
with the model-independent value derived from the $QCD$-based
heavy-quark expansion, $|V_{ub} / V_{cb}| = 0.098 \pm 0.013$
\cite{NU96}, and with the value extracted from the measurement of
the end-point region of the lepton spectrum, $|V_{ub} / V_{cb}| =
0.08 \pm 0.01_{exp} \pm 0.02_{th}$ \cite{CLEO93,ARGUS90}, as well
as with the result obtained from a $LF$ analysis of the exclusive
decays $B \to D \ell \nu_{\ell}$ and $B \to \pi \ell \nu_{\ell}$,
$|V_{ub} / V_{cb}| = 0.082 \pm 0.016$ \cite{SIM96}. Furthermore,
our result for $|V_{ub}|$ (Eq. (\ref{4.3})) is consistent within 
the errors with the value $|V_{ub}| = (2.9 \pm 0.4) \cdot 10^{-3}$,
obtained in Ref. \cite{SIM96} from a $LF$ analysis of the exclusive
decay $B \to \pi \ell \nu_{\ell}$, and with the finding $|V_{ub}| =
(3.2 \pm 0.4) \cdot 10^{-3}$, obtained in Ref. \cite{DKND97} after
averaging over the exclusive $B \to \pi \ell \nu_{\ell}$ and $B
\to \rho \ell \nu_{\ell}$ decay modes, as well as with the result
$|V_{ub}| = (3.3 \pm 0.2_{-0.4}^{+0.3} \pm 0.5) \cdot 10^{-3}$
quoted in a recent $CLEO$ report \cite{Berk96}.

\indent  Note that the larger uncertainty in our extracted value of
$|V_{ub}|$ (Eq. (\ref{4.3})) is the experimental one, mainly
because of the large error quoted by the $ALEPH$ collaboration
\cite{ALEPH}. An interesting way to obtain a better determination
of $|V_{ub}|$ has been proposed recently in Ref. \cite{BDU97} and
it is based on the investigation of the recoil mass spectrum in
the $B \to X_u \ell \nu_{\ell}$ decays at $M_X \le M_{max} < M_D$.
The suggested value for $M_{max}$ is $1.5 ~ GeV$, chosen in order
to avoid the leakage of the tail of the $B \to X_c \ell
\nu_{\ell}$ transitions, which can occur because of the finite
resolution of the experiments. We have therefore evaluated the
partial branching ratio, $\Gamma_{SL}^* / \Gamma_B^{(exp)}$,
obtained by cutting the integration over $s$ in Eq. (\ref{3.11})
at the value $s_{max} = (1.5 ~ GeV / m_b)^2$. The results for
$\Gamma_{SL}^* / \Gamma_B^{(exp)}$ in units of $10^{-2} \cdot
(|V_{ub}| / 0.0032)^2 \cdot (\tau_B^{(exp)} / 1.57 ~ ps)$ are
$0.073, ~ 0.064, ~ 0.074, ~ 0.078, ~ 0.060$ in case of the quark
models $A - E$, respectively, which correspond to a fraction of $b
\to u$ decays with $M_X \le 1.5 ~ GeV$ equal to $67 \%$, $59 \%$,
$61 \%$, $63 \%$ and $59 \%$. Their average is $0.070 \pm 0.008$
in comparison with the total (averaged) branching ratio of $0.113
\pm 0.009$ in the same units (cf. Table 3), corresponding to a
fraction of $(62 \pm 4) \%$ of $b \to u$ decays with $M_X \le 1.5
~ GeV$. It follows that the model dependence is only slightly
increased by cutting the integration over the recoil mass $M_X$ at
$M_{max} = 1.5 ~ GeV$, at the price of reducing the number of $B
\to X_u \ell \nu_{\ell}$ events by $\simeq 40 \%$, in overall
agreement with the findings of Ref. \cite{BDU97}.

\section{Conclusions}

\indent In this paper we have derived a new parton formula, which
establishes a simple connection between the decay rate of a free
heavy-quark and the one of a heavy-quark bound in a hadron. The main
features of our approach are the treatment of the $b$-quark as an
on-mass-shell particle and the inclusion of the effects arising from
the $b$-quark transverse motion in the $B$-meson. Our main result is
Eq. (\ref{1.2}), or more precisely Eqs. (\ref{3.11}-\ref{3.12}).

\indent Another result of this paper is the determination of the
$CKM$ parameters $|V_{cb}|$ and $|V_{ub}|$ using the available
experimental values for the branching ratio of the processes $B \to
X_c \ell \nu_{\ell}$ and $B \to X_u \ell \nu_{\ell}$. Our results
exhibit a model dependence related mainly to the uncertainties
associated to the non precise knowledge of the primordial $b$-quark
distribution function and to the values of the constituent quark
masses. These uncertainties lead to a final theoretical uncertainty
in the extracted value of both $|V_{cb}|$ and $|V_{ub}|$ of about
$\pm 5 \%$. Including leading-order  perturbative $QCD$ corrections,
we have found $|V_{cb}| = (43.0 \pm 0.7_{exp} \pm 1.8_{th}) \cdot
10^{-3}$ and $|V_{ub}| = (3.83 \pm 0.48_{exp} \pm 0.14_{th}) \cdot
10^{-3}$, which imply $|V_{ub} / V_{cb}| = 0.089 \pm 0.011_{exp} \pm
0.005_{th}$, in nice agreement with existing predictions.

\indent In conclusion, we point out that our $LF$ approach can be
applied to the investigation of lepton energy spectra and to
inclusive processes other than semileptonic decays, like, e.g., the
non-leptonic branching ratios both for external and internal types
of decays \cite{MTM96,GNST96}. Another field of application is the
calculation of the lifetime of the $B_c$ meson and the theoretical
estimation of the inclusive $b \to s \gamma$ width.

\section*{Acknowledgments}

\indent Three authors (S.Ya.K., I.M.N. and K.A.T.-M.) gratefully
aknowledge the financial support of the INTAS-RFBR grant, ref. No
95-1300 and the INTAS grant, ref. No 96-195. This work was in part
supported by the RFBR grant, ref. No 95--02--04808a and RFBR
grant, ref. No 96--15--96740.

\newpage

\section*{Appendix}

\indent Let us remind that the dimensionless tensor $W_{\alpha
\beta}$ in Eq. (\ref{2.5}) can be decomposed into five different
Lorentz covariants, involving therefore five structure functions,
$W_1(q^2, q_0)$ to $W_5(q^2, q_0)$, viz:
 \be
    \label{A.1}
    W_{\alpha \beta} & = & -g_{\alpha \beta} W_1(q^2, q_0) +
    v_{\alpha} v_{\beta} W_2(q^2, q_0) - i\epsilon_{\alpha \beta
    \gamma \delta} u^{\gamma} v^{\delta} W_3(q^2, q_0) +
    \nonumber \\
    & & (v_{\alpha} u_{\beta} + u_{\alpha} v_{\beta}) W_4(q^2, q_0)
    + u_{\alpha} u_{\beta }W_5(q^2, q_0),
 \ee
where $v \equiv P_B / M_B$ is the four-velocity of the $B$-meson
and $u \equiv q / M_B$. Note that the terms proportional to
$u_{\alpha}$ or $u_{\beta}$ in Eq. (\ref{2.7}) do not contribute
to the differential decay rate if the lepton masses can be
neglected. Therefore, for $\ell = e$ or $\mu$ only $W_1(q^2,
q_0)$, $W_2(q^2, q_0)$ and $W_3(q^2, q_0)$ are actually relevant.
The dimensionless structure functions $W_i(q^2, q_0)$ are
functions of the two invariants $q^2$ and $q_0 \equiv v \cdot q$.

\indent All the inclusive observables can be expressed in terms of
the structure functions $W_i(q^2, q_0)$ ($i = 1, 5$). Thus, the
contraction $<L^{\alpha\beta}> W_{\alpha\beta} / M_B^2$ in
Eq. (\ref{2.8}) is given by
 \be
    \label{A.2}
    {<L^{\alpha \beta}> W_{\alpha \beta} \over M_B^2} = {4 \over 3}
    ~ x_0^2 ~ F(t, s), 
 \ee
where
 \be
    \label{A.3}
    F(t, s) & \equiv & 3t (1 + \lambda_1 - 2\lambda_2) ~ W_1(t,
    s) + \left[ (1 + \lambda_1) {|\vec{q}|^2 \over m_b^2} + {3
    \over 2} \lambda_2 t \right] ~ W_2(t, s) + \nonumber \\
    & & {3 \over 2} \lambda_2 t \left[ (1 + x_0^2 t - x_0^2 s) ~
    W_4(t, s) + x_0^2 t ~ W_5(t, s) \right] .
 \ee

We have derived our basic result (\ref{3.11}-\ref{3.12}) without
calculating explicitly the structure functions $W_i(t, s)$ in our
$LF$ parton approximation (\ref{3.1}). However, since we have in
mind also more general applications, like, e.g., the calculation
of the lepton energy spectrum, we now present our $LF$ formulae
for the functions $W_i(t, s)$. To this end, after averaging over
the azimutal angle ${\varphi}$ of the transverse momentum
$\vec{p}_{\perp}$, the quark tensor $w_{\alpha \beta}^{bq'}(p_b,
p_{q'})$ can be cast into the form
 \be
    \label{A.4}
    \int {d\varphi \over 2\pi} ~ w_{\alpha \beta}^{(bq')}(p_b,
    p_{q'}) & = & m_b^2 \left\{ -g_{\alpha \beta} w_1 + v_{\alpha}
    v_{\beta} w_2 - i\epsilon_{\alpha \beta \gamma \delta}
    v^{\gamma} u^{\delta} w_3 + \right. \nonumber \\
    & & \left. (v_{\alpha} u_{\beta} + v_{\beta} u_{\alpha}) w_4 +
    u_{\alpha} u_{\beta} w_5 \right\},
\ee
which implies ($i = 1, 5$)
 \be
    \label{A.5}
    W_i(t, s) = m_b^2 {\pi \over r} \int\limits_{x_1}^{min[1,
    x_2]} dx ~ w_i(t, s; x) ~ |\psi(x, p_{\perp}^2)|^2 .
 \ee
where $r \equiv x_0 ~ q^+ / m_b$. Note that the contraction
$<L^{\alpha \beta}> w^{bq'}_{\alpha \beta}$ of the lepton and
quark tensors is given in terms of the functions $w_i$ by
 \be
    \label{A.6}
    {<L^{\alpha \beta}> w^{bq'}_{\alpha \beta} \over m_b^4} = {4
    \over 3} \tilde{F}(t,\rho), 
 \ee
where
 \be
    \label{A.7}
    \tilde F(t, \rho) & \equiv & 3t (1 + \lambda_1 - 2\lambda_2) ~
    w_1 + \left[ (1 + \lambda_1) {|\vec{q}|^2 \over m_b^2} + {3
    \over 2} \lambda_2 t \right] ~ w_2 + \nonumber \\
    & & {3 \over 2}\lambda_2 t \left[ (1 + x_0^2 t - x_0^2 s) ~ w_4
    + x_0^2 t ~ w_5 \right ]. 
 \ee
The calculation of the functions $w_i(t, s; x)$ is straightforward
and leads to
 \be
    \label{A.8}
    w_1(t, s; x) & = & 2 [(\xi - 1/2) z_0 - \xi^2 t],
    \nonumber  \\
    w_2(t, s; x) & = & {8 \over \alpha^2(t, s)} ~ \left[ 3\xi^2 t^2
    - 3\xi t z_0 + t + {1 \over 2} z_0^2 \right],
    \nonumber \\
    w_3(t, s; x) & = & - {2 \over x_0 \alpha(t, s)} \left[ z_0 -
    2\xi t \right], \nonumber \\
    w_4(t, s; x) & = & {4 \over x_0^2 \alpha^2(t, s)} \left\{
    \left[ x - {x_0^2 \over r} (z_0 - \xi t) - {x_0 \alpha(t, s)
    \over 2} \right] \left( z_0 - 2\xi t  \right) + \right.
    \nonumber \\
    & & \left. [\xi z_0 - \xi^2 t -1] (1 + x_0^2 t -
    x_0^2 s) \right\}, \nonumber \\
    w_5(t, s; x) & = & {4 \over x_0^4 \alpha^2(t, s)} \left\{
    \left[ x - {x_0^2 \over r} (z_0 - \xi t) \right] ~ 
    \left[ x - x_0^2 {\xi \over x} (z_0 - \xi t) -
    \right. \right. \nonumber \\
    & & \left. \left. x_0 \alpha(t, s) \right] - 2x_0^2 \left[ \xi
    z_0 - \xi^2 t -1 \right] \right\},
 \ee
where $\xi \equiv x / r$, $z_0 \equiv 1 + t - \rho$ and the
quantity $\alpha(t,s)$ is defined by Eq. (\ref{2.9}). Using Eq.
(\ref{A.8}) it can be easily verified that
 \be
   \label{A.9}
   \tilde{F}(t, \rho) = (1 + \lambda_1) [(1 - \rho)^2 + (1 + \rho)
   t - 2t^2] - 3\lambda_2 t(1 + \rho - t).
 \ee
Using Eqs. (\ref{3.9}), (\ref{A.6}) and (\ref{A.9}) one obtains the
well-known result
 \be
    \label{A.10}
    {d\Gamma^{(free)}_{SL}(t) \over dt} & = & {G_F^2 m_b^5 \over
    48\pi^3} ~ |V_{bq'}|^2 ~ {|\vec{\tilde{q}}|  \over m_b} ~
    \Phi_{\ell}(t) ~ \left \{ (1 + \lambda_1) [(1 - \rho)^2 + (1 +
    \rho) t - 2t^2] - \right. \\ \nonumber
    & & \left. 3\lambda_2 t(1 + \rho - t) \right \}.
 \ee

\newpage

\noindent {\bf Table 1}. The values of the constituent quark masses
(in $GeV$) and of the average internal momentum squared $<p^2>$
(in $GeV^2$) for the quark models $A - E$ (see text). The value
of the B-meson mass is $M_B = 5.279 ~ GeV$ \cite{PDG96}.

\vspace{0.25cm}

\begin{center}

\begin{tabular}{|c||c|c|c|c|c|}
\hline
Model & A & B & C & D & E \\
\hline \hline
$m_b$                  & 4.800 & 4.880 & 5.200 & 5.237 & 4.977 \\
\hline
$m_c$                  & 1.400 & 1.550 & 1.820 & 1.835 & 1.628 \\
\hline
$m_u$                  & 0.300 & 0.330 & 0.330 & 0.337 & 0.220 \\
\hline \hline
$M_B - m_b$            & 0.479 & 0.399 & 0.079 & 0.042 & 0.302 \\
\hline
$m_b - m_c$            & 3.400 & 3.330 & 3.380 & 3.402 & 3.349 \\
\hline
$m_b - m_u$            & 4.500 & 4.550 & 4.870 & 4.900 & 4.757 \\
\hline \hline
$m_b / M_B$            & 0.909 & 0.924 & 0.985 & 0.992 & 0.943 \\
\hline
$m_c / m_b$            & 0.292 & 0.318 & 0.350 & 0.350 & 0.327 \\
\hline
$m_u / m_b$            & 0.063 & 0.068 & 0.063 & 0.064 & 0.044 \\
\hline \hline
$<p^2>$                & 0.234 & 0.252 & 0.277 & 0.262 & 0.553 \\
\hline
\end{tabular}

\end{center}

\vspace*{1cm}

\noindent {\bf Table 2}. The branching ratio $\Gamma_{SL} /
\Gamma_B^{(exp)}$ for the inclusive process $B \to X_c \ell
\nu_{\ell}$ (with $\ell = e, \mu$) in units of $10^{-2} \cdot
(|V_{cb}| / 0.040)^2 \cdot (\tau_B^{(exp)} / 1.57 ~ ps)$,
calculated within the quark models $A - E$ and considering for 
the $pQCD$ corrections an overall reduction factor equal to 
$0.90$ \cite{pQCD}. The values of
$|V_{cb}|$ in units of $10^{-3} \cdot \sqrt{Br_{SL}^{(exp)} /
10.43 ~ \%} \cdot \sqrt{1.57 ~ ps / \tau_B^{(exp)}}$, obtained
within the various quark models considered, are also reported.

\vspace{0.25cm}

\begin{center}

\begin{tabular}{|c||c|c|c|c|c|}
\hline
Model & A & B & C & D & E \\
\hline \hline
$\Gamma_{SL} / \Gamma_B^{(exp)}$     & 10.1 & ~9.2 & ~9.0 &
                                       ~9.1 & ~8.0 \\ \hline
\hline
$|V_{cb}|$                           & 40.7 & 42.6 & 43.1 &
                                       42.8 & 45.6 \\ \hline
\end{tabular}

\end{center}

\vspace*{1cm}

\noindent {\bf Table 3}. The branching ratio $\Gamma_{SL}
/ \Gamma_B^{(exp)}$ of the inclusive process $B \to X_u \ell
\nu_{\ell}$ (with $\ell = e, \mu$) in units $10^{-2} \cdot
(|V_{ub}| / 0.0032)^2 \cdot (\tau_B^{(exp)} / 1.57 ~ ps)$,
calculated within the quark models $A - E$ and considering for 
the $pQCD$ corrections an overall reduction factor equal to 
$0.85$ \cite{pQCD}. The values of
$|V_{ub}|$ in units of $10^{-3} \cdot \sqrt{Br_{SL}^{(exp)} / 0.16
~ \%} \cdot \sqrt{1.57 ~ ps / \tau_B^{(exp)}}$ and the values of
the ratio $|V_{ub} / V_{cb}|$, obtained within the various quark
models $A - E$, are also reported.

\vspace{0.25cm}

\begin{center}

\begin{tabular}{|c||c|c|c|c|c|}
\hline
Model & A & B & C & D & E \\
\hline \hline
$\Gamma_{SL} / \Gamma_B^{(exp)}$ & 0.109 & 0.108 & 0.121 & 0.123 &
                                   0.102 \\ \hline \hline
$|V_{ub}|$                       & 3.88~ & 3.89~ & 3.69~ & 3.64~ &
                                   4.01~ \\ \hline
$|V_{ub} / V_{cb}|$              & 0.095 & 0.091 & 0.086 & 0.085 &
                                   0.088 \\ \hline
\end{tabular}

\end{center}

\end{document}